# Coexistence of Ferroelectric and Relaxor-like Phases in a Multiferroic Solid Solution $(1-x)Pb(Fe_{1/2}Nb_{1/2})O_3 - xPbMnO_3$

Anna N. Morozovska[1*], Victor N. Pavlikov[2], Yuriy O. Zagorodniy[2], Iryna V. Kondakova[2], Oleksandr S. Pylypchuk[1], Andrii V. Bodnaruk[1], Oksana V. Leshchenko[2], Myroslav V. Karpets[2,3], Roman O. Kuzian[2], and Eugene A. Eliseev[2†]

[1] *Institute of Physics, National Academy of Sciences of Ukraine, 46, Nauky Avenue, 03028 Kyiv, Ukraine*

[2] *Frantsevich Institute for Problems in Materials Science, National Academy of Sciences of Ukraine, 3, str. Omeliana Pritsaka, 03142 Kyiv, Ukraine*

[3] *Ye. O. Paton Institute of Materials Science and Welding, National Technical University of Ukraine "Igor Sikorsky Kyiv Polytechnic Institute", 37, Beresteisky Avenue, Kyiv, Ukraine, 03056*

## Abstract

Experimental and theoretical studies of unusual polar, dielectric and magnetic properties of room temperature multiferroics, such as perovskites $Pb(Fe_{1/2}Nb_{1/2})O_3$ (PFN) and $Pb(Fe_{1/2}Ta_{1/2})O_3$ (PFT), are very important. We study the phase composition, dielectric, ferroic properties of the solid solutions PFN and PFT substituted with 5, 10, 15, 20 and 30 % of Mn ions prepared by the solid-state synthesis. The XRD analysis confirmed the perovskite structure of sintered ceramics. Electric measurements revealed the ferroelectric-type hysteresis of electric charge in pure PFN ceramics and in PFN ceramics substituted with (10 – 30)% of Mn. At the same time, the PFN−5% Mn ceramics did not show any ferroelectric properties due to very high conductivity. Magnetostatic measurements reveal the ferromagnetic properties of PFN−5 % Mn ceramics, and a paramagnetic behavior or a very weak antiferromagnetic-like saturation of magnetization in other PFN−Mn and PFT−Mn ceramics. Temperature dependences of the dielectric permittivity of PFN−10% Mn and PFN−15% Mn ceramics have two pronounced maxima, one of which is relatively sharp and has a weak frequency dispersion; another is diffuse and has a strong frequency dispersion. A further increase in the Mn content up to 20% leads to the right shift in the paraelectric-ferroelectric phase transition temperature, as well as to the strong suppression of the second wide maximum, which transforms into a small diffuse shoulder. An increase in the Mn substitution up to 30% leads to a significant decrease in the dielectric permittivity, left shift of its maximum, and induces a pronounced frequency dispersion of the paraelectric-ferroelectric transition temperature, which is inherent to relaxor-like ferroelectrics. Thus, the appearance of ferroelectricity, magnetic and dielectric properties depend significantly and non-monotonically on the concentration of Mn ions. To explain observed dependencies, we evolved a theoretical model describing unusual polar and dielectric properties of PFN–Mn and PFT–Mn ceramics by introducing the Edwards-Anderson order parameter. Comparison of the model with experiments reveal the coexistence of the ordered ferroelectric-like and disordered relaxor-like phases in the multiferroic solid solutions PFN–Mn and PFT–Mn.

---

[*] Corresponding author: anna.n.morozovska@gmail.com
[†] Corresponding author: eugene.a.eliseev@gmail.com

## I. INTRODUCTION

A strong magnetoelectric (ME) coupling existing at room temperature is especially important for fundamental science and novel functional devices fabrication [1, 2, 3, 4, 5]. Promising examples are the solid solutions of ferroelectric antiferromagnets $Pb(Fe_{1/2}Ta_{1/2})O_3$ (PFT) and $Pb(Fe_{1/2}Nb_{1/2})O_3$ (PFN), which attract much attention [6, 7, 8, 9].

For PFN the antiferromagnetic Neel transition temperature $T_N$ and ferroelectric Curie temperature $T_C$ are $T_N$ = 143–170 K [10, 11] and $T_C$ = 379–393 K [12, 13], while $T_N$ = 133–180 K [14, 15] and $T_C \cong 250$ K [16] for PFT. PFN is an antiferromagnet with G-type spin ordering below at $T<T_{Neel}$, where the Neel temperature $T_{Neel}$ = 143 - 170 K. PFN is a ferroelectric at temperatures $T<T_{Curie}$, where its Curie temperature varies in the range $T_{Curie}$ = 379 – 393K. PFN has a biquadrartic ME coupling constant $2.2 \times 10^{-22}$ sm/(VA) at 140K. PFT is an antiferromagnet with the Neel temperature $T_{Neel}$ = 133 - 180 K and ferroelectric at $T<T_{Curie}$ with the Curie temperature $T_{Curie} \approx 250$ K. PFT has a biquadrartic ME coupling constant of the same order as that of PFN. Saturated and low loss ferroelectric hysteresis curves with a remanent polarization of about 20-30 $\mu C/cm^2$ was observed in Refs. [17].

Note that the theoretical consideration [18] had shown that the giant linear ME coefficient is also possible due to the size effect in nanostructured PFT–PZT lamellas. Later, Landau-Ginzburg-Devonshire (LGD) thermodynamic formalism was proposed for the description of the anomalous ferroelectric, ferromagnetic and ME properties of $Pb(Fe_{1/2}Ta_{1/2})_x(Zr_{0.53}Ti_{0.47})_{1-x}O_3$ and $Pb(Fe_{1/2}Nb_{1/2})_x(Zr_{0.53}Ti_{0.47})_{1-x}O_3$ micro-ceramics [19].

In this work we study the phase composition, magnetic and ferroelectric of the PFN and PFT solid solutions with 0, 5, 10, 15 and 30 % of Mn ions, further denoted as PFN-Mn and PFT–Mn, respectively. The X-ray diffraction (XRD) analysis confirms the perovskite structure of the sintered solid solutions. Electric measurements reveal ferroelectric-type hysteresis in the PFN–Mn and PFT–Mn ceramics. Magnetostatic measurements reveal the ferromagnetic properties of PFN–10 %Mn, meanwhile other PFN-Mn and PFT-Mn ceramics demonstrate paramagnetic behavior or a very weak antiferromagnetic saturation of magnetization. We evolved a theoretical model focused on the multiferroic properties and phase coexistence in the PFN–Mn and PFT–Mn ceramics. Comparison of the model with experiments reveal the coexistence of the ordered ferroelectric-like and disordered relaxor-like phases in the multiferroic solid solutions PFN–Mn and PFT–Mn.

## II. FERROIC PROPERTIES OF PFN AND PFT WITH MN IONS
### A. Experimental Results

The solid solutions, PFN–Mn and PFT–Mn, were prepared by solid-state synthesis in air at 800°C for 6 hours, followed by calcination at 1000 – 1200°C for 4 and 2 hours (see details in **Supplement S1**). In this way we sintered thee types of solid solutions in the form of ceramic samples:

Type I. $(1 - x)\text{Pb}(\text{Fe}_{0.5}Me_{0,5})\text{O}_3 - x\text{PbMnO}_3$, $Me$ = Nb, Ta; the content x = 5 – 30 mol.%.

Type II. $(1 - x)\text{Pb}(\text{Fe}_{0.5}Me_{0,5})\text{O}_3 - x\text{Mn}_2\text{O}_3$, $Me$ = Nb, Ta; where the substitution degree "x" of $\text{Fe}_2\text{O}_3$ by $\text{Mn}_2\text{O}_3$ was 5, 10, 15, 20, and 30 mol. %.

Type III. Pure $\text{Pb}(\text{Fe}_{0.5}Me_{0,5})\text{O}_3$, $Me$ = Nb, Ta, for comparison with Mn-substituted samples.

The XRD analysis confirmed the perovskite structure of the sintered solid solutions (see **Figs. S1-S2**). For example, the XRD pattern of a PFN–15 % Mn, consists of the 99.4 wt.% perovskite phase after calcination at 800°C for 6 hours. To measure electrical properties, the samples were prepared in the form of round tablets with a diameter $D \approx 7$ mm and a thickness $d \approx 1.5$ mm. Silver electrodes were deposited om their planar surface. The measurements of charge-voltage and current-voltage characteristics, capacitance and loss tangent were carried out in the flat capacitor geometry. The values of the dielectric permittivity were calculated using the expression for the flat capacitor, $\varepsilon = \frac{Cd}{\varepsilon_0 \pi D^2}$.

Electric measurements reveal the ferroelectric-type hysteresis of electric charge in PFN ceramics (see the black loop in **Fig. 1**), as well as in PFN – 10 % Mn and PFN – 15 % Mn ceramics (see the red and magenta loops in **Fig. 1**). At that the polarization hysteresis loop becomes significantly lower ($P_r$ decreases) and wider ($E_c$ increases) in the PFN – 15 % Mn ceramics. A high electroconductivity leads to the blowing of the polarization hysteresis loops in PFN – 20 % Mn and PFN – 30 % Mn (see the blue and green loops in **Fig. 1**). However, the PFN–5% Mn ceramics does not have any ferroelectric-like properties, which we relate to its very high conductivity. At the same time, magnetostatic measurements revealed the pronounced ferromagnetic properties of PFN – 5 % Mn ceramics, meanwhile all other PFN–Mn and PFT–Mn samples demonstrate either the paramagnetic behavior or a very weak antiferromagnetic saturation of magnetization (see **Fig. S3** in **Supplement S2** for details).

Thus, the appearance and the shape of the ferroelectric-like hysteresis loops, the magnitude of remanent charge density $Q_r$ and coercive field $E_c$ depend non-monotonically on the content "x" of Mn ions (compare different loops in **Fig. 1**). Considering the contribution of free carriers, we concluded that the remanent polarization $P_r$, which can be extracted from the charge-voltage loops, decreases slightly with increase in Mn content "x" in PFN – x Mn ceramics. Note that the electric conductivity depends on the preparation conditions and can be different for samples with different "x".

The observed effects can be explained in the following way. The influence of Mn-substitution on $P_r$ and $E_c$ is conditioned by the local disorder in the PFN induced by Mn ions. In result the solid solution can be considered as a mixture of regions with increased concentration of Fe ions, which have the properties of an ordered ferroelectric, and the regions with increased concentration of Mn ions, which can reveal the properties of a disordered relaxor-like ferroelectric. A relatively high conductivity allows us to suspect the spatial coexistence of the ordered and disordered phases, possibly separated by semiconducting boundaries.

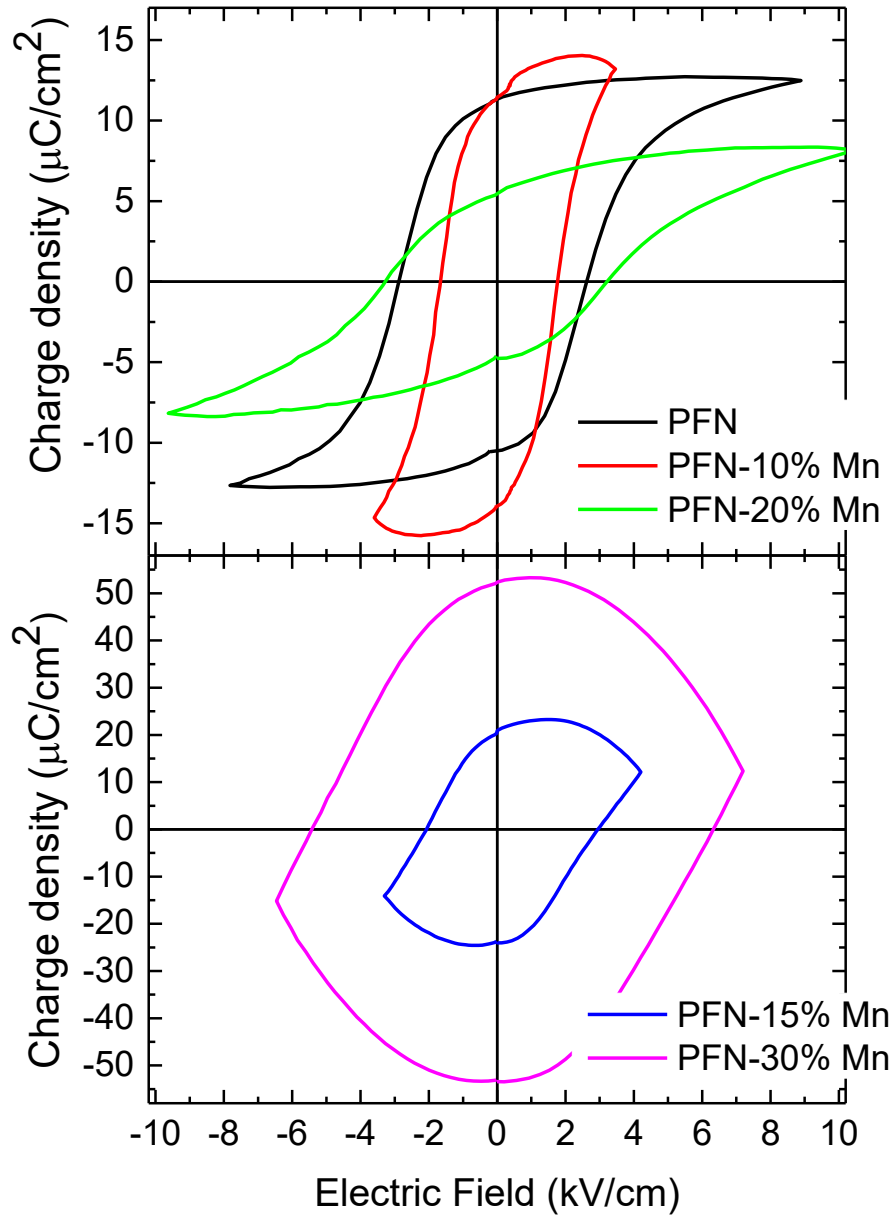

**Figure 1.** Charge density hysteresis field dependence of the $(1 - x)Pb(Fe_{0.5}Nb_{0.5})O_3 - xPbMnO_3$ ceramics with x = 0, 10, 15, 20 and 30 mol. %.

The dielectric measurements were performed in the temperature range from 30 to 275°C using an immittance meter E7-20. Temperature dependences of the dielectric permittivity and losses measured for PFN-Mn ceramics are shown in **Fig. 2**. The permittivity curves depend on the Mn content in a rather nontrivial way, which may be explained using the following argumentation.

PFN is a compositionally disordered perovskite ferroelectric with a diffuse paraelectric-ferroelectric phase transition with $T_C \cong 106 - 120°C$ [12, 13]. It exhibits relaxor properties, but its permittivity does not have a visible dependence of the phase transition temperature on the measurement frequency (see e.g., Refs. [10, 20]). This is usually explained by the random distribution of Nb and Fe atoms at the B sites of the perovskite structure due to the approximately equal radius of the $Fe^{3+}$ and $Nb^{5+}$ ions (0.78 Å). Substitution of $Fe^{3+}$ by $Mn^{4+}$ with a substantially smaller ionic radius (0.67 Å) leads to lattice deformation, as well as to the formation of various vacancies with different valence states. The temperature dependences of the dielectric permittivity of $Pb(Fe_{1/2}Nb_{1/2})O_3$ – 10 % Mn and $Pb(Fe_{1/2}Nb_{1/2})O_3$ – 15 % Mn have two peaks (see **Figs. 2(a) and 2(b)**). The position of the first one (near 115°C) is close to the PE-FE phase transition temperature of PFN and the peak position slightly depends on frequency. The second peak is much wider and located near 200°C; it has a pronounced frequency dispersion, and its intensity decreases very strongly with increase in frequency. The position of the second peak coincides with the start of a sharp increase in dielectric losses, as clearly seen in **Figs. 2(a) and 2(b)**. This is likely due to the activation of charge carriers and the formation of space charges, leading to a Maxwell-Wagner-type polarization associated with the presence of vacancies, which increase the conductivity of ceramic grains. At low frequencies, the contribution of much less conductive grain boundaries is significant, more conductive grains contribute with increase in frequency, thereby reducing the dielectric permittivity. These Maxwell-Wagner-type effects stipulate enhanced interfacial polarization dynamics, leading to the interfacial barrier-layer capacitance (IBLC) effect [21], as well as to the inhomogeneous layers between the electrodes and the ceramic surface, known as the surface barrier layer capacitance (SBLC) effect [22]. Thus, large dielectric permittivity (~40000) observed in the PFN-xMn samples near 200°C can be related to the synergy of IBLC and SBLC effects at the interior of insulating grain boundaries [21, 22, 23], and/or to polaron hopping in semiconducting grain cores with a large number of vacancies (see e.g., Refs. [24, 25]).

A further increase in the manganese content up to 20% leads to the right shift in the PE-FE phase transition temperature to 125°C, as well as to the strong suppression of the second wide maximum, which transforms into a small diffuse shoulder (see **Figs. 2(c)**). An increase in the manganese substitution up to 30% leads to a significant decrease in the maximal permittivity (below 20000), shift to the left its position (near 90 – 100 °C) and induces a pronounced frequency dispersion of the PE-FE transition temperature, which is inherent to relaxor-like ferroelectrics (see **Fig. 2(d)**).

The temperature dependence of the permittivity in ferroelectric with diffuse phase transition can be described by the modified Curie-Weiss law, $\varepsilon(T) \sim \frac{C_W}{[(T-T_C)^2+\Delta_d^2(\omega)]^{\frac{v}{2}}}$, where the parameter $v$ can vary from 1 for ordered ferroelectrics to 2 for relaxor-like disordered ferroelectrics [26], $\Delta_d^2(\omega)$ is the frequency dependent dispersion modelling the diffuse phase transition. Fitting the experimental data yields $v \approx 1.6$ for PFN – 10 % Mn and PFN – 15 % Mn ceramics; and $v \approx 1.4$ for PFN – 20 % Mn ceramics. At the same time $v \approx 1.9$ for PFN – 30 % Mn ceramics, which is close to values for relaxor ferroelectrics. The increase of $v$ is apparently due to the formation of a larger number of vacancies, which is confirmed by the increase in the conductivity of the PFN – 30 % Mn sample.

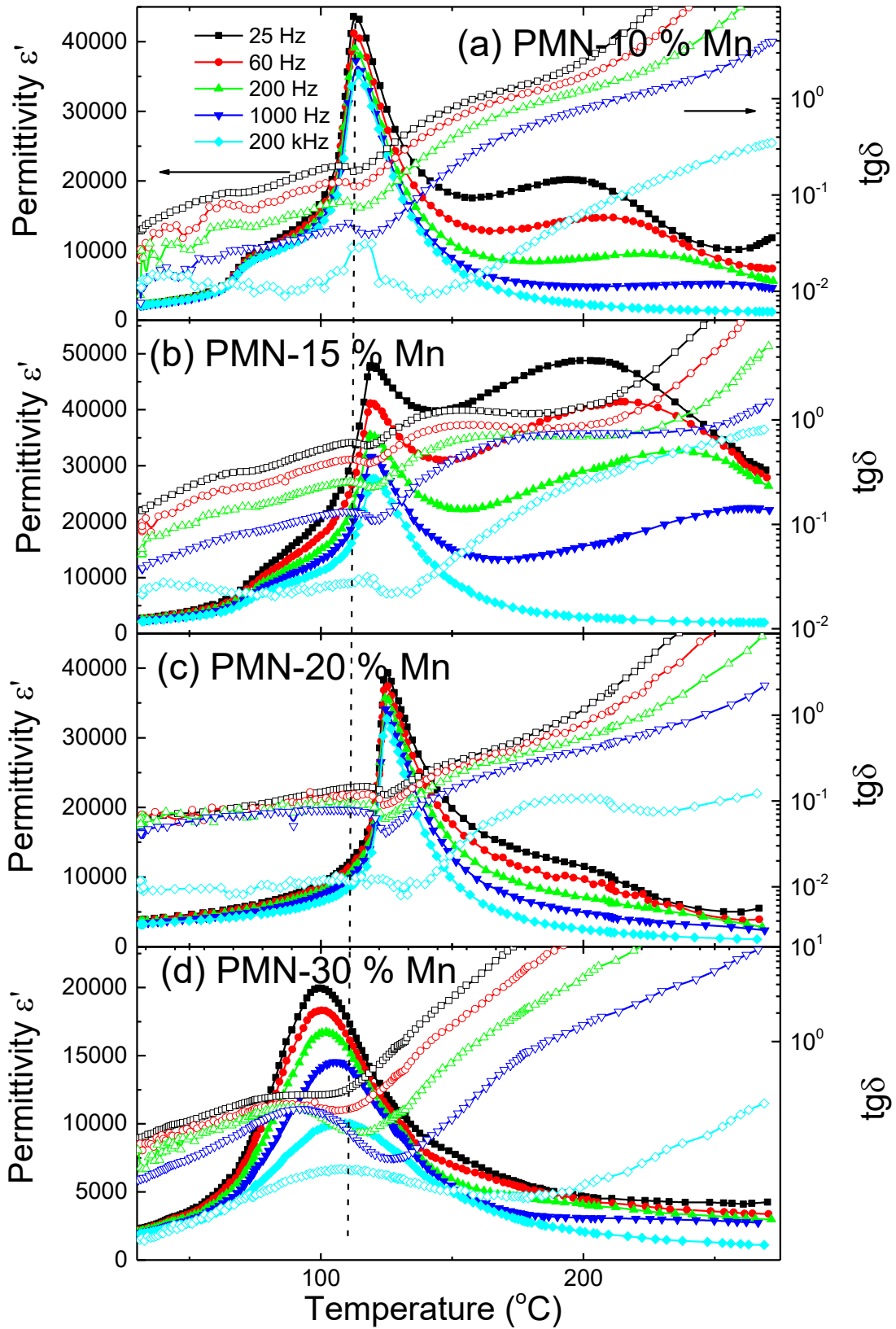

**Figure 2.** Temperature dependence of the effective dielectric permittivity of the (1 − x)Pb(Fe$_{0.5}$Nb$_{0.5}$)O$_3$ – xPbMnO$_3$ ceramics with x = 10, 15, 20 and 30 mol. %. Different curves correspond to the frequences 25 Hz, 60 Hz, 2 kHz and 200 kHz.

## B. Microscopic Explanation

The dielectric and magnetic properties of PFN and PFT are highly sensitivity to the chemical disorder at the B-sites, which are randomly occupied by $Fe^{3+}$ and $Nb^{5+}$ ($Ta^{5+}$) ions, because their ionic radii and bonding preferences are nearly identical. In the ideal ordered perovskite lattice, $Pb^{2+}$ shows a stereoactive $6s^2$ lone pair, occurring due to the hybridization between the Pb 6$s$ and O 2$p$ orbitals, contributes to the ferroelectric (FE) distortion and can compensate, to a certain extent, the suppression of the FE distortion caused by incorporation of magnetic ions with a partially filled $d$-shell.

In disordered PFN (PFT), the local charge randomness from $Fe^{3+}/Nb^{5+}$ ($Ta^{5+}$) substitution suppresses coherent Pb–O displacements, yielding diffuse maxima of dielectric susceptibility instead of sharp peak in a chemically ordered $Pb(Fe_{1/2}Sb_{1/2})O_3$ as shown in Ref. [27]. That is, disorder leads to a nanoscale distribution of polar regions, similar to $PbMg_{1/3}Nb_{2/3}O_3$ (PMN). Occasionally, nanosized Fe-rich regions can form, but they lack long-range periodicity. $Fe^{3+}$ ions (S = 5/2) interact through Fe–O–Fe superexchange ($J_1 \approx$ 50–70 K). Random positioning of Fe and Nb breaks the percolation of magnetic bonds, producing a spin-glass state below 25 K.

Partial replacement $Fe^{3+} \rightarrow Mn^{3+}/Mn^{4+}$ introduces mixed-valence centers. Based on the results for $PbMnO_3$ [28], it can be predicted that Mn in PFN will form mixed states of $Mn^{3+}/Mn^{4+}$ and cause a local charge disproportion $Pb^{2+} \rightarrow Pb^{4+}$, which changes the electronic structure and polarization potential. The local charge transfer between Pb and Mn can occur, perturbing the Pb 6s–O 2p hybridization responsible for ferroelectricity. The change in the valence of Mn creates additional donor/acceptor centers that screen the Pb–O electric dipoles and the maximum $\varepsilon'(T)$ decreases. Mn has a smaller ionic radius than other B-site lattice ions and causes local strains. This effect further broadens the relaxation temperature distribution, i.e. makes the system a "deeper relaxor". At the same time, Mn–O octahedra distort more strongly than Fe–O ones, creating additional local dipoles and possibly enhancing low-T dielectric permittivity. At low concentrations (~5–10%), Mn can increase the polaron conductivity, so an additional conductive contribution is superimposed on $\varepsilon'(T)$, which gives the peak asymmetry.

## III. THEORETICAL DESCRIPTION
## A. Landau-Ginzburg-Devonshire Approach

The bulk density of the LGD free energy, corresponding to the PFN-Mn or PFT-Mn, is the sum of polarization energy ($g_P$), antiferromagnetic ($g_L$) and ferromagnetic ($g_M$) energies, elastic energy ($g_{el}$), antiferrodistortive energy ($g_{AFD}$) and magnetoelectric energy ($g_{ME}$) [19, 29]:

$$G_{LGD} = g_P + g_L + g_M + g_{el} + g_{AFD} + g_{ME}, \quad (1)$$

The contributions to Eq. (1) are listed in **Supplement S3**. The linear and quadratic ME energies include AFD-FE, AFD-FM and AFD-AFM couplings [29]. Variation of the free energy (1) via the FE, FM, AFM and AFD order parameters yields the time-dependent LGD-Khalatnikov equations of state [19], which are also listed in **Supplement S3**.

In what follows we assume that the AFD long-range order significantly affects the FE, AFM and FM long-range orders, but not vice versa, as argued in most previous works [9, 19, 30, 31]. Also, the FE long-range order significantly affects the AFM and FM long-range orders, but not vice versa, for the same reasons. Using the above arguments and the percolation theory approach [32], the antiferromagnetic-paramagnetic (AFM-PM), ferromagnetic-paramagnetic (FM-PM) and ferroelectric-paraelectric (FE-PE) transition temperatures can be estimated similarly to that was done in Ref. [19] (see details in **Supplement S3**).

According to the LGD approach, the temperature dependence of the dielectric permittivity $\varepsilon_C(T, x, \omega)$ can be fitted by the following function near the diffuse ferroelectric-like maximum [33, 34]:

$$\varepsilon_C(T, x, \omega) = \varepsilon_b^0(x) + \left( \frac{C_W(x)}{\sqrt{\left(T - T_C(x) + 3\tilde{\beta}T_C(x)p^2(x,T) + 5\tilde{\gamma}T_C(x)p^4(x,T)\right)^2 + \Delta_d^2(x,\omega)}} \right)^\nu. \quad (2)$$

Here $\varepsilon_b^0(x)$ is the non-ferroelectric "background" permittivity, which is a temperature-independent constant or very weak linear function of temperature. The positive constant $C_W(x)$ is the analog of the Curie-Weiss constant, $T_C(x)$ is the Curie temperature of the diffuse ferroelectric transition, $\Delta_d^2(x, \omega) = \langle E_i^r(x, \vec{r}) E_i^r(x, \vec{r}') \rangle$ is the permittivity dispersion determined by the dispersion of random electric field $E_i^r(x, \vec{r})$. The condition $\Delta_d(x, \omega) \ll T_C$ corresponds to a relatively sharp transition at $T = T_C$. The dimensionless fitting parameters $\tilde{\beta}$ and $\tilde{\gamma}$ can vary in the range $-1 < \tilde{\beta}$ and $\tilde{\gamma} \geq 0$. The power $v$ varies from 1 (for ordered displacive ferroelectrics with the second order phase transition) to 2 (for disordered ferroelectrics) [26].

For an ordered ferroelectric with the first (or second) order PE-FE phase transition the dimensionless polarization $p(x, T)$ can be introduced as

$$p(x,T) = \sqrt{\frac{2|1-(T/T_C(x))|}{\tilde{\beta}+\sqrt{\tilde{\beta}^2+4|1-(T/T_C(x))|\tilde{\gamma}}}} \frac{1}{1+\exp\left[\frac{T-T_C(x)}{\Delta_d(x,\omega)}\right]}. \quad (3)$$

The parameter $\tilde{\beta} < 0$ for the first order PE-FE phase transition, and $\tilde{\beta} > 0$ for the second order PE-FE phase transition. The dispersion $\Delta_d(x, \omega)$ models the diffuseness of the polar order parameter transition to the ordered state. The diffuseness of the transition is related to the frozen disorder due to the random electric fields.

For relaxor (or relaxor-like) ferroelectrics the temperature dependence of the polar order parameter $p(x,T)$ is under debate. Rather, it should be substituted by the Edwards-Anderson (EA) order parameter, $q_{EA}(x,T)$, inherent to disordered systems and spin/dipolar glasses (see e.g., Ref.[35] and refs. therein). The EA order parameter quantifies the degree of frozen local dipoles (or correlated order) in a disordered relaxor ferroelectric. The EA parameter zero in the disordered PE phase and becomes non-zero below the PE-FE phase transition temperature, indicating that the system disordered spin-glass state. In this sense, the proportionality $q_{EA}(x,T) \sim \frac{1}{1+\exp\left[\frac{T-T_f(x)}{\Delta_f(x,\omega)}\right]}$ maybe still valid, but the "freezing" temperature $T_f(x)$ can be different from the aristo-phase Curie temperature $T_C(x)$, as well as the dispersion $\Delta_f(x, \omega)$ can be different from the permittivity dispersion $\Delta_d(x, \omega)$. The spherical random bond-random theory predicts the following dependence for dielectric permittivity of the disordered ferroelectrics in dipolar glass state (see e.g. Ref. [36]):

$$\varepsilon_C(T,x) = \varepsilon_b^0(x) + \frac{C_W(x)\{1-q_{EA}(x,T)\}}{T-T_C(x)\{1-q_{EA}(x,T)\}}, \quad (4)$$

where $q_{EA}(x,T)$ is the EA order parameter.

## B. Comparison with experiment

Expression for the dielectric permittivity can be generalized in the case of mixture considered in the effective medium approximations (shortly "EMA") [37, 38, 39]. Most of these approximations are applicable for quasi-spherical randomly distributed dielectric (or semiconducting) particles in the insulating environment, such as the Lichtenecker-Rother approximation of the logarithmic mixture [40]. As a rule, EMA considers the algebraic equation for the effective permittivity $\varepsilon^*_{eff}(T, x, \omega)$ of the binary mixture [33]:

$$(1-\mu)\frac{\varepsilon^*_{eff}-\varepsilon^*_b}{(1-n_a)\varepsilon^*_{eff}+n_a\varepsilon^*_b} + \mu\frac{\varepsilon^*_{eff}-\varepsilon^*_a}{(1-n_a)\varepsilon^*_{eff}+n_a\varepsilon^*_a} = 0. \quad (5)$$

Here $\varepsilon^*_a$, $\varepsilon^*_b$ are complex functions of the relative permittivity of the components "a" and "b" respectively, $\mu$ and $1-\mu$ are relative volume fractions of the components "a" and "b" respectively, $n_a$ is the depolarization field factor for the inclusions of the type "a". In the case $\mu=0$ or $\mu=1$, the solution of Eq.(4) is $\varepsilon^*_{eff} = \varepsilon^*_b$ or $\varepsilon^*_{eff} = \varepsilon^*_a$, respectively.

For $n_a = 1$ (i.e., for the system consisting of the layers, perpendicular to the external field) the solution of Eq.(5) gives the expression for the Maxwell layered dielectric model, $\varepsilon^*_{eff} = \left(\frac{1-\mu}{\varepsilon^*_b} + \frac{\mu}{\varepsilon^*_a}\right)^{-1}$. For $n_a = 1/3$ (i.e., for the system consisting of the spherical nanoparticles) the solution of Eq.(6) is $\varepsilon^*_{eff} = \varepsilon^*_b\left[1 + \frac{3\mu(\varepsilon^*_a-\varepsilon^*_b)}{\varepsilon^*_a+2\varepsilon^*_b-\mu(\varepsilon^*_a-\varepsilon^*_b)}\right]$, which is the Wagner expression. For $n_a = 0$ (i.e., for the system of the columns, parallel to the external field) Eq.(5) yields $\varepsilon^*_{eff} = (1-\mu)\varepsilon^*_b + \mu\varepsilon^*_a$, which is equivalent to the system with parallel connected capacitors with the complex permittivity $\varepsilon^*_b$ and $\varepsilon^*_a$.

To describe the unusual dielectric properties of PFN-Mn, we use the following fitting functions:

$$\varepsilon^*_a(T, x, \omega) = \varepsilon^0_b(x) + \varepsilon_C(T, x, \omega) - i\frac{\sigma_a(T)}{\omega}, \quad (6a)$$

$$\varepsilon^*_b(T, x, \omega) = \varepsilon^0_b(x) + \frac{C_m(x)}{(T-T_m)^2 + \Delta^2_b(x,\omega)} - i\frac{\sigma_b(T)}{\omega}. \quad (6b)$$

Here $\varepsilon_C(T, x, \omega)$ is given by Eqs.(2) or (4), $\sigma_a$ and $\sigma_b$ are the conductivities of components "a" and "b".

Expressions (5)-(6) allow to fit the effective dielectric permittivity considering IBLC and SBLC effects. As a starting point we put $C_m(x) = 0$, $n_a = 1/3$ and $\sigma_a \gg \sigma_b$, which correspond to spherical ferroelectric-semiconducting grains in the non-ferroelectric environment with much smaller conductivity (a classical Maxwell-Wagner case).

Temperature dependences of the effective dielectric permittivity of the (1 – x)Pb(Fe$_{0.5}$Nb$_{0.5}$)O$_3$ – xPbMnO$_3$ ceramics calculated for x = 10, 15, 20, and 30 mol. % are shown in **Figs. 3(a) - 3(d)**, respectively. Empty symbols are experimental results measured at 200 kHz, when the contribution of the Maxwell-Wagner effects is small enough; and filled symbols are experimental results measured at 25 Hz, when the contribution of the Maxwell-Wagner effects dominates above $T_C(x)$.

Black dashed curves in **Figs. 3** are theoretical fittings of empty symbols using the disordered model (2) for $\varepsilon'_a$ with $\nu > 1.0$ and other parameters listed in the second column of **Table I.** It is seen that the parameter $\nu$ varies from 1.2 to 1.3, which is quite reasonable. However, the effective Curie-Weiss constant $C_W$ varies from $0.212 \cdot 10^5$ K to $1.234 \cdot 10^5$ K, which is a very large scattering without a clear physical reason. The scattering of Curie temperature is also quite large and nonmonotonic. Thus, we should conclude that the disordered model (2) does not give us quantitative information about the polar state of ceramic grains, while it confirms the relaxor-like features.

Black solid lines in **Figs. 3** are theoretical fittings of empty symbols using EA model (4) for $\varepsilon'_a$ with $\nu = 1.0$ and other parameters listed in the third column of **Table I.** Temperature dependences of the EA order parameter $q_{EA}(T)$, deconvoluted from the effective dielectric permittivity of (1 – x)Pb(Fe$_{0.5}$Nb$_{0.5}$)O$_3$ – xPbMnO$_3$ ceramics, are shown in **Fig. 4**. It is typical for relaxor ferroelectrics [36]. The proximity of red, blue, green and magenta curves calculated for x = 10, 15, 20, and 30 mol. %, respectively, confirms the relaxor-like state of the ceramic grains entire x-range. Therefore, we can conclude that the EA model (4) gives us quantitative information about the polar state of ceramic grains.

It should be underlined that the temperatures $T_f(x)$ and $T_C(x)$ can be separated for x = 10 and 15 mol. %, indicating the normal ferroelectric-like and relaxor-like phases coexistence in the range of Mn content. For the x = 30 mol.% the relaxor state dominates.

Significant deviations from the Curie-Weiss law are observed at low frequencies (see filled symbols at 25 Hz) and/or higher temperatures (above 150°C), indicating the domination of the Maxwell-Wagner effects. However, even in this case we may suspect the relaxor-like state in the intergrain.

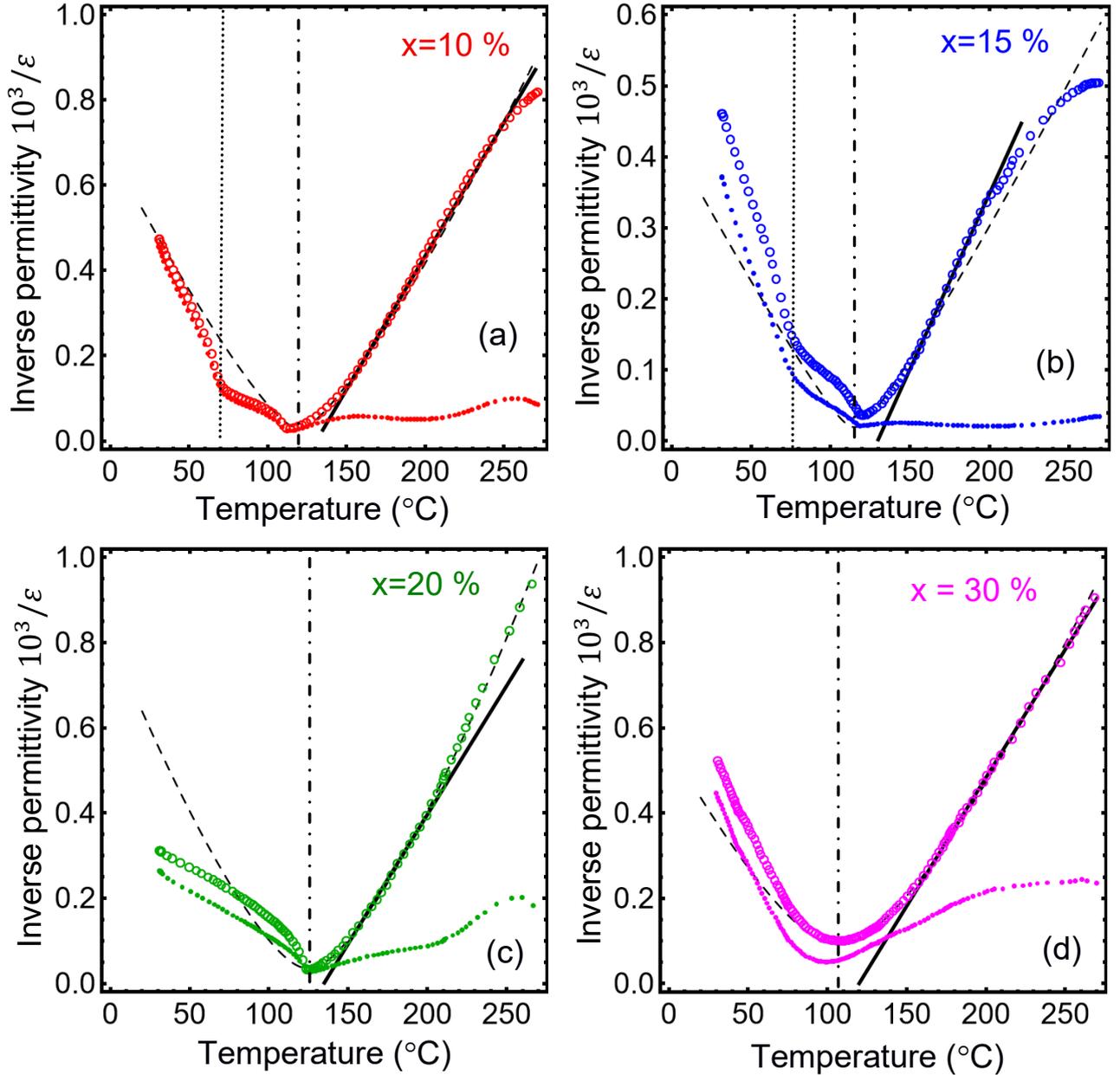

**Figure 3.** Temperature dependences of the inverse dielectric permittivity $1/\varepsilon_{eff}$ of the $(1-x)Pb(Fe_{0.5}Nb_{0.5})O_3 - xPbMnO_3$ ceramics calculated for x = 10 (the red curve), 15 (the blue curve), 20 (the green curve), and 30 mol. % (the magenta curve). Empty and filled symbols are experimental results measured at 200 kHz and 25 Hz, respectively. Black solid and dashed

curves are a theoretical fitting by Eqs.(2)-(6) with parameters listed in **Table I**. Dotted and dash-dotted vertical lines point on the temperatures $T_f(x)$ and $T_C(x)$, respectively.

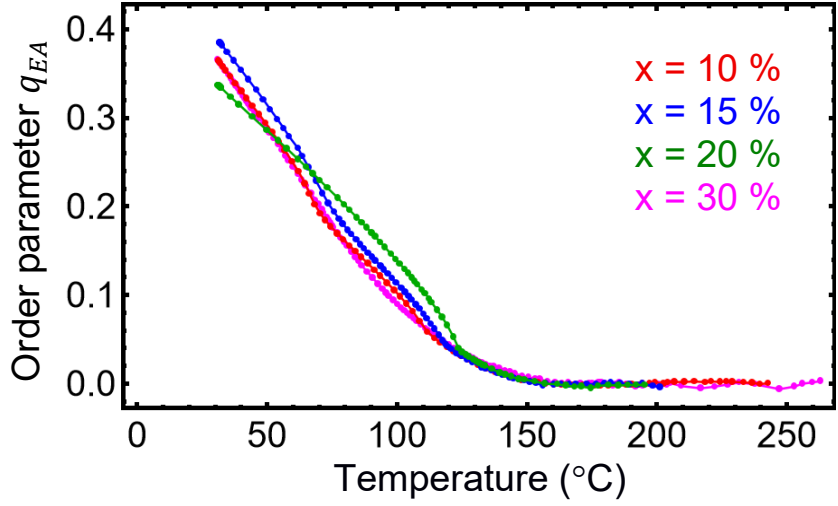

**Figure 4.** Temperature dependences of the EA order parameter $q_{EA}(T)$ deconvoluted from the dielectric permittivity of the $(1 - x)Pb(Fe_{0.5}Nb_{0.5})O_3 - xPbMnO_3$ ceramics with x = 10 (the red curve), 15 (the blue curve), 20 (the green curve), and 30 mol. % (the magenta curve).

TABLE I. The values of the fitting parameters used to describe the effective dielectric permittivity of the $(1 - x)Pb(Fe_{0.5}Nb_{0.5})O_3 - xPbMnO_3$ ceramics at frequency $f = 200$ kHz.

| Content x (%) and fraction $\mu$ | Model based on Eqs.(2), (5) and (6) | Model based on Eqs.(2), (5) and (6) |
|---|---|---|
| x = 10 %, $\mu = 0.10$ | $\varepsilon_b^0 = 7$, $\Delta_d = 9.4$ K, $C_W = 0.461 \cdot 10^5$ K, $\nu = 1.23$, $T_C = 120°C$ | $\varepsilon_b^0 = 7$, $C_W = 1.567 \cdot 10^5$ K, $T_C = 130.9°C$, $\nu = 1.0$ |
| x = 15 %, $\mu = 0.15$ | $\varepsilon_b^0 = 7$, $\Delta_d = 7.3$ K, $C_W = 1.234 \cdot 10^5$ K, $\nu = 1.11$, $T_C = 115°C$ | $\varepsilon_b^0 = 7$, $C_W = 2.017 \cdot 10^5$ K, $T_C = 129.6°C$, $\nu = 1.0$ |
| x = 20 %, $\mu = 0.20$ | $\varepsilon_b^0 = 7$, $\Delta_d = 13.0$ K, $C_W = 0.212 \cdot 10^5$ K, $\nu = 1.39$, $T_C = 125°C$ | $\varepsilon_b^0 = 7$, $C_W = 1.641 \; 10^5$ K, $T_C = 134.8°C$, $\nu = 1.0$ |
| x = 30 %, $\mu = 0.30$ | $\varepsilon_b^0 = 7$, $\Delta_d = 27.6$ K, $C_W = 0.361 \cdot 10^5$ K, $\nu = 1.29$, $T_C = 107°C$ | $\varepsilon_b^0 = 7$, $C_W = 1.651 \cdot 10^5$ K, $T_C = 120.0°C$, $\nu = 1.0$ |

## IV. CONCLUSIONS

We study the phase composition, dielectric, ferroelectric and magnetic properties of the solid solutions PFN and PFT substituted with 5, 10, 15, 20 and 30 % of Mn ions prepared by the solid-state synthesis. The XRD analysis confirmed the perovskite structure of sintered ceramics. Magnetostatic measurements reveal the ferromagnetic properties of PFN - 10 %

Mn, and the paramagnetic behavior or a very weak antiferromagnetic-like saturation of magnetization in other PFN-Mn and PFT-Mn ceramics.

Electric measurements revealed the ferroelectric-type hysteresis of electric charge in pure PFN ceramics and in PFN ceramics substituted with (10 – 30)% of Mn. At the same time, the PFN–5% Mn ceramics did not show any ferroelectric properties due to very high conductivity. Magnetostatic measurements reveal the ferromagnetic properties of PFN–5 % Mn ceramics, and a paramagnetic behavior or a very weak antiferromagnetic-like saturation of magnetization in other PFN–Mn and PFT–Mn ceramics.

Temperature dependences of the PFN–10% Mn and PFN–15% Mn ceramics dielectric permittivity have two pronounced maxima, one of which (located near 125°C) is relatively sharp and has a weak frequency dispersion; another (located near 200°C) is diffuse and has a very strong frequency dispersion. A further increase in the Mn content up to 20% leads to the right shift in the paraelectric-ferroelectric phase transition temperature, as well as to the strong suppression of the second wide maximum, which transforms into a small diffuse shoulder. An increase in the Mn substitution up to 30% leads to a significant decrease in the dielectric permittivity, left shift of its maximum, and induces a pronounced frequency dispersion of the paraelectric-ferroelectric transition temperature, which is inherent to relaxor-like ferroelectrics.

Thus, the appearance of ferroelectricity, magnetic and dielectric properties depend significantly and non-monotonically on the concentration of Mn ions. To explain observed dependencies, we evolved a theoretical model describing the polar and dielectric properties of the PFN-Mn and PFT-Mn ceramics. The influence of Mn-substitution on the PFN-Mn and PFT-Mn solid solutions is conditioned by the local disorder induced by Mn ions. In result the PFN-Mn and PFT-Mn solid solutions can be considered as a mixture of regions with increased concentration of Fe ions, which have the properties of an ordered ferroelectric, and the regions with increased concentration of Mn ions, which can reveal the properties of a disordered relaxor-like ferroelectric with a corresponding EA order parameter. A relatively high conductivity allows us to suspect the spatial coexistence of the ordered and disordered phases, possibly separated by semiconducting boundaries. Comparison of the model with experiments confirm the coexistence of the ordered ferroelectric-like and disordered relaxor-like phases in the multiferroic solid solutions PFN-Mn and PFT-Mn.

## Acknowledgments

The work of A.N.M. is funded by the NAS of Ukraine, the Target Program of the National Academy of Sciences of Ukraine, Project No. 5.8/25-П "Energy-saving and environmentally friendly nanoscale ferroics for the development of sensorics, nanoelectronics and spintronics". A.V.B. are funded by the NAS of Ukraine, grant No. 07/01-2025(6) "Nano-sized multiferroics with improved magnetocaloric properties". The work of O.S.P. is funded by the Ministry of Science and Education of Ukraine, contract M/35-2025 "Flexible nano-ferroelectrics for rapid cooling of combat electronics". The work of I.V.K., Y.O.Z. V.N.P., OV.L. and E.A.E is funded by the NAS of Ukraine.

## SUPPLEMENT

### Supplement S1. Samples preparation and XRD characterization

The starting materials were oxides (>99.5 %) PbO, $Fe_2O_3$, $Nb_2O_5$, $Ta_2O_5$, and MnAc. After decomposition of the MnAc in the mixtures at 500°C, the powders were ground in a planetary mill for 6 hours with $ZrO_2$ balls in distilled water, dried and pressed into disk-shaped samples 10 mm in diameter and 2 mm in height. The developed mixture preparation technology proved effective already at the synthesis stage. In this way, two types of solid solutions, namely PFN-Mn and PFT-Mn, in the form ceramic samples were sintered by solid-phase synthesis in air at 800°C for 6 hours, followed by calcination at temperatures of 1000 – 1200°C for 4 and 2 hours. These two types are:

(1) $(1 - x)Pb(Fe_{0.5}Me_{0.5})^{4+}O_3 - xPbMn^{4+}O_3$, $Me$ = Nb, Ta; the content x = 5 – 30 mol.%;

(2) $Pb(Fe_{0.5}Me_{0.5})^{4+}O_3 - Mn_2O_3$, $Me$ = Nb, Ta; substitution of $Fe_2O_3$ by $Mn_2O_3$ was carried out in quantities of 5, 10, 15, 20, 30 and 40 mol. %, based on its content in the formula compositions $Pb(Fe_{1/2}Nb_{1/2})O_3$ (PFN) and $Pb(Fe_{1/2}Ta_{1/2})O_3$ (PFT).

**Figure S1** shows the X-ray diffraction pattern of a PFN-based solid solution with 15 mol.% of $PbMnO_3$, where the perovskite phase content is 99.4 wt.% after calcination at 800°C for 6 hours.

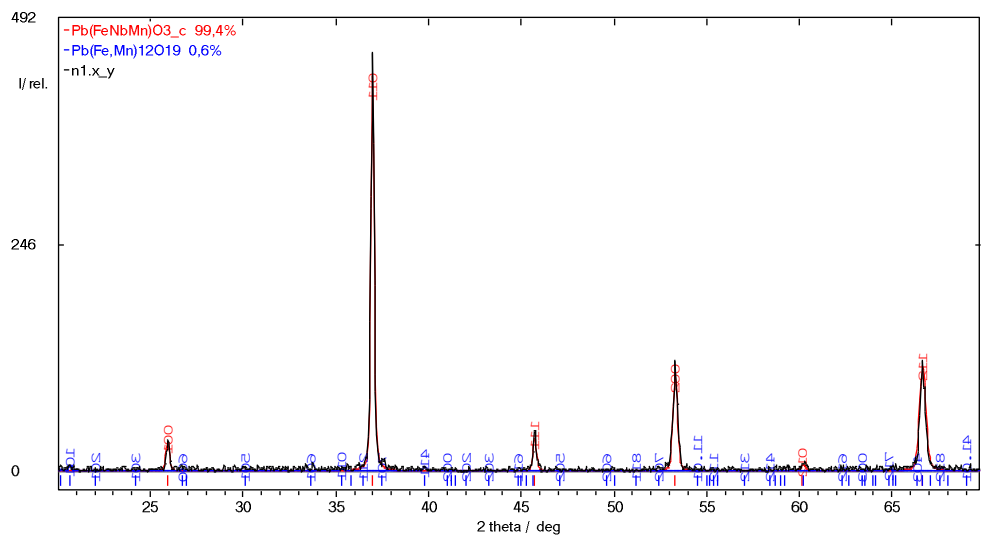

**Figure S1.** Diffraction pattern of a sample of composition 85PFN- 15PbMn$_2$O$_3$ after firing in air at 800C for 6 hours.

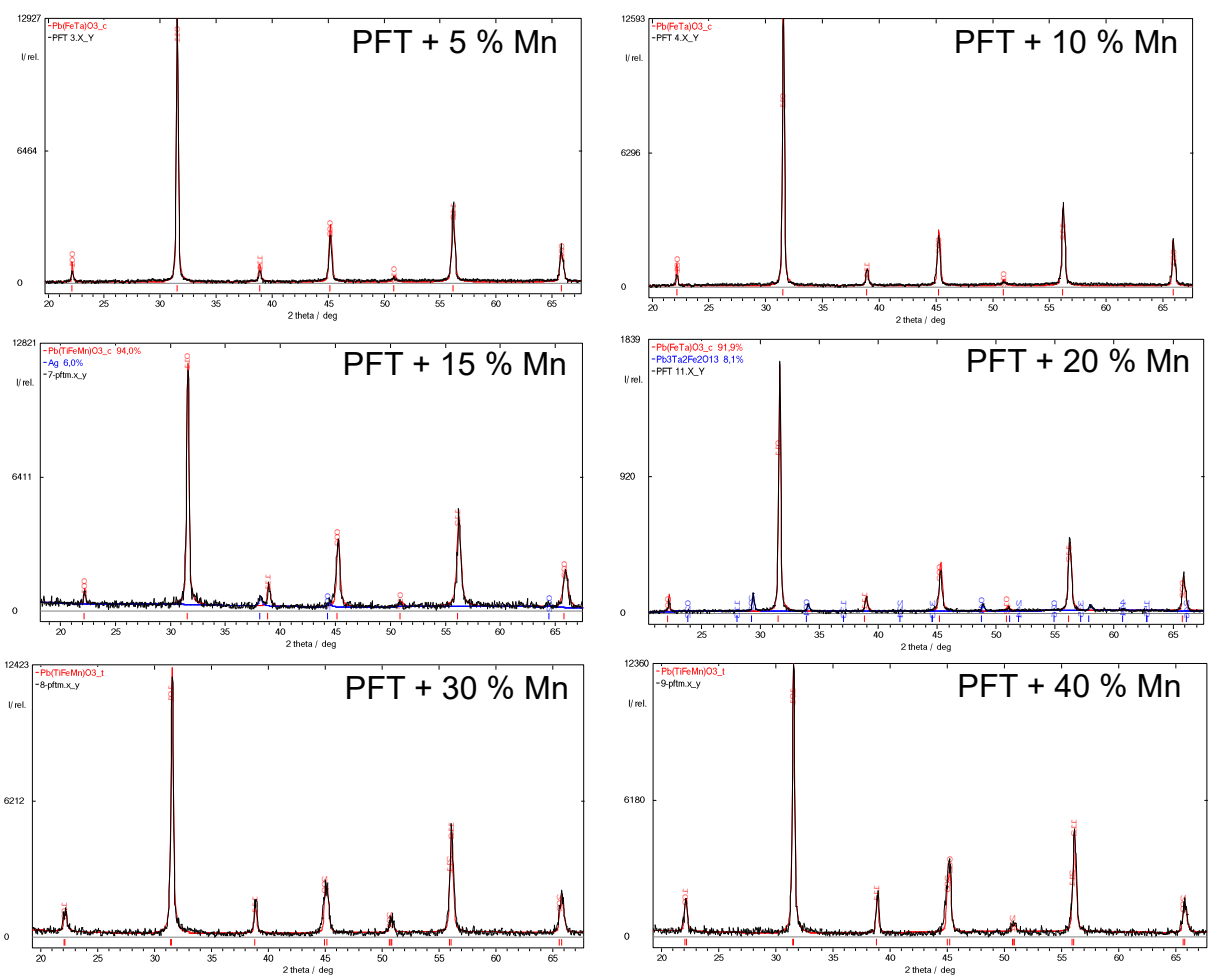

**Figure S2.** Typical XRD spectra of PFT substituted with 0, 5, 10, 15, 20, 30 and 40 % of Mn ions.

## Supplement S2. Magnetostatic measurements

Magnetostatic measurements, performed using an LDJ 9500 magnetometer with a vibrating sample, reveal the pronounced ferromagnetic properties of PFN – 5 % Mn (see orange curve in **Fig. S3(a)**). Other PFN–Mn and PFT–Mn samples, with x = 0, 10, 15, 20 and 30 % of Mn ions, demonstrate either the paramagnetic behavior (for x = 0) or a very weak antiferromagnetic saturation of magnetization (for x = 10, 15, 20 and 30 % of Mn) (see e.g., red, black, green, blue and magenta curves in **Fig. S3(a)** and all curves in **Fig. S3(b)**).

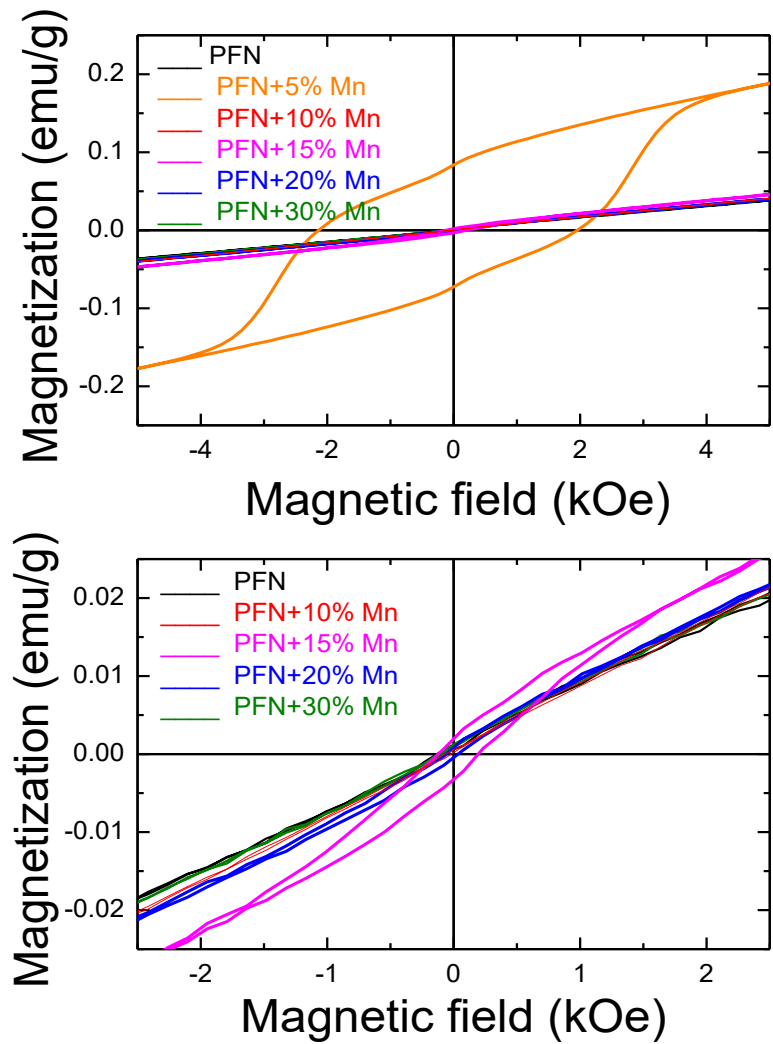

(a)

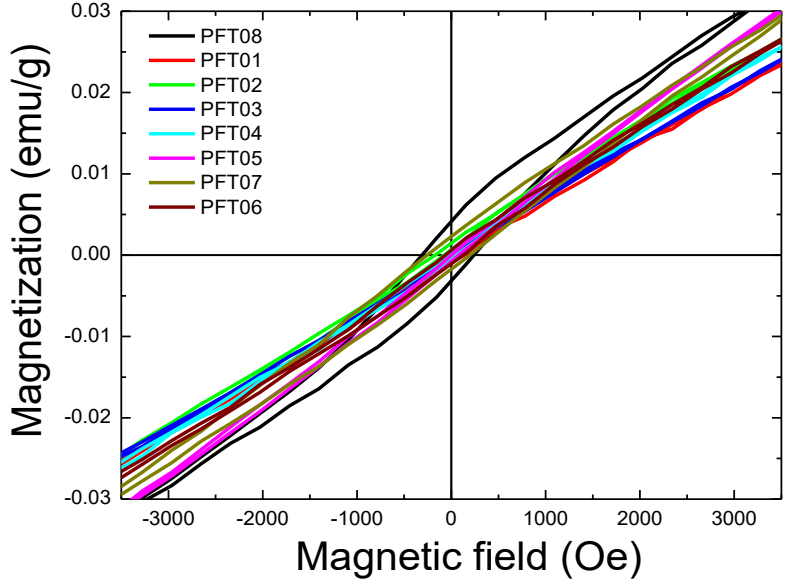

(b)

**Figure S3.** Magnetization field dependence of the **(a)** $(1 - x)Pb(Fe_{0.5}Nb_{0.5})O_3 - xPbMnO_3$ ceramics with x = 0, 10, 15, 20, 30 mol. % and **(b)** PFT substituted with 0, 5, 10, 15, 20, 30 and 40 % of Mn ions, denoted as PFT01-08. Substitution of $[PbFeTa_{0.5}O_5]^{4+}$ by $Mn^{4+}$ in $PbTa_{0.5}Fe_{0.5}O_3$ takes place in the PFT01-PFT05 samples with 5, 10, 15, 20 and 30 % of Mn ions. Substitution of $Fe^{+3}$ by $Mn^{3+}$ in $PbTa_{0.5}Fe_{0.5}O_3$ takes place in the PFT06-PFT08 samples with 10, 30 and 40 % of Mn ions.

## Supplement S3. Landau-Ginzburg-Devonshire Approach

The bulk density of the LGD free energy, corresponding to the PFN-Mn or PFT-Mn, is the sum of polarization energy ($g_P$), antiferromagnetic ($g_L$) and ferromagnetic ($g_M$) energies, elastic energy ($g_{el}$), antiferrodistortive energy ($g_{AFD}$) and magnetoelectric energy ($g_{ME}$):

$$G_{PM} = g_P + g_L + g_M + g_{el} + g_{AFD} + g_{ME}, \qquad (S.1a)$$

$$g_P = \frac{\alpha_P}{2} P_i^2 + q_{ijkl}^{(e)} u_{ij} P_k P_l + \frac{\beta_{Pij}}{4} P_i^2 P_j^2 + \frac{\gamma_{Pijkl}}{6} P_i^2 P_j^2 P_k^2 P_l^2 - E_i^r(x, \vec{r}) P_i, \qquad (S.1b)$$

$$g_L = \frac{\alpha_L}{2} L_i^2 + q_{ijkl}^{(l)} u_{ij} L_k L_l + \frac{\beta_{Lij}}{4} L_i^2 L_j^2 + \frac{\gamma_{Lijkl}}{6} L_i^2 L_j^2 L_k^2 L_l^2, \qquad (S.1c)$$

$$g_M = \frac{\alpha_M}{2} M_i^2 + q_{ijkl}^{(m)} u_{ij} M_k M + \frac{\beta_{Mij}}{4} M_i^2 M_j^2 + \frac{\gamma_{Lijkl}}{6} M_i^2 M_j^2 M_k^2 M_l^2 - H_i^r(x, \vec{r}) M_i, \qquad (S.1d)$$

$$g_{el} = \frac{c_{ijkl}}{2} u_{ij} u_{kl} + \frac{A_{ijklmn}}{2} u_{ij} u_{kl} P_m P_n + \frac{B_{ijklmn}}{2} u_{ij} u_{kl} L_m L_n + \frac{C_{ijklmn}}{2} u_{ij} u_{kl} M_m M_n, \qquad (S.1e)$$

$$g_{AFD} = \frac{\alpha_\Phi}{2} \Phi_i^2 + q_{ijkl}^{(r)} u_{ij} \Phi_k \Phi_l + \frac{\beta_{\Phi ij}}{4} \Phi_i^2 \Phi_j^2. \qquad (S.1f)$$

In equations (S.1) $P_k$ are the components of ferroelectric (FE) polarization, $L_i = (M_{ai} - M_{bi})/2$ and $M_i = (M_{ai} + M_{bi})/2$ are the components of antiferromagnetic (AFM) and ferromagnetic (FM) order parameters, respectively. They correspond to the two equivalent sub-lattices $a$ and $b$, respectively. $u_{ij}$ are the components of elastic strain tensor. $\Phi_i$ are the components of the antiferrodistortive (AFD) order parameter. $c_{ijkl}$ are the components of elastic stiffness. $q_{ijkl}^{(e)}$, $q_{ijkl}^{(l)}$, $q_{ijkl}^{(r)}$ and $q_{ijkl}^{(m)}$ are the bulk electrostriction, antimagnetostriction, rotostriction and magnetostriction tensor components, respectively. $E_i^r(x, \vec{r})$ and $H_i^r(x, \vec{r})$ are the random electric and magnetic fields induced by the presence of Mn ions and vacancies.

The coefficients $\alpha_P$, $\alpha_L$, $\alpha_M$ and $\alpha_\Phi$ depend linearly on temperature, namely $\alpha_P = \alpha_{TY}(T - T_Y^C)$, $\alpha_L = \alpha_{LT}(T - T_Y^N)$, $\alpha_M = \alpha_{MT}(T - \theta_Y^C)$ and $\alpha_\Phi = \alpha_{\Phi T}(T - T_Y^\Phi)$, where Y="N" or "T"; "N" corresponds to PFN and "T" corresponds to PFT. $T_Y^C$ is the ferroelectric Curie temperature of a bulk material; $T_Y^N$ and $\theta_Y^C$ are antiferromagnetic Néel temperature and ferromagnetic Curie temperature, respectively; $\alpha_\Phi = \alpha_{\Phi T}(T - T_Y^\Phi)$, $T_Y^\Phi$ is the antiferrodistortive transition temperature. The LGD coefficients, the temperatures $T_Y^N$, $T_Y^\Phi$ and $\theta_Y^C$ should depend on the content "x" of Mn ions.

The linear and quadratic ME energies include AFD-FE, AFD-FM and AFD-AFM couplings:

$$g_{ME} = \begin{pmatrix} \mu_{ij} M_i P_j + \tilde{\mu}_{ij} L_i P_j + \frac{\eta_{ijkl}^{FM}}{2} M_i M_j P_k P_l + \frac{\eta_{ijkl}^{AFM}}{2} L_i L_j P_k P_l + \\ \frac{\eta_{ijkl}^{\Phi P}}{2} \Phi_i \Phi_j P_k P_l + \frac{\eta_{ijkl}^{\Phi M}}{2} \Phi_i \Phi_j M_k M_l + \frac{\eta_{ijkl}^{\Phi L}}{2} \Phi_i \Phi_j L_k L_l \end{pmatrix}. \quad (S.2)$$

In Equation (S.2), $\mu_{ij}$ are the components of the bilinear ME coupling tensor, $\eta_{ijkl}^{FM}$ and $\eta_{ijkl}^{AFM}$ are the components of the biquadratic ME coupling tensor, $\eta_{ijkl}^{\Phi P}$, $\eta_{ijkl}^{\Phi M}$ and $\eta_{ijkl}^{\Phi L}$ are the components of the AFD-FE, AFD-FM and AFD-AFM coupling tensors, respectively. The ME coupling coefficients should depend on the content "x" of Mn ions.

Variation of the free energy (S.1) via FE, FM, AFM and AFD order parameters yields the time-dependent LGD-Khalatnikov equations of state:

$$\Gamma_P \frac{\partial P_i}{\partial t} + \left( \alpha_P \delta_{il} - \eta_{mnil}^{FM} M_m M_n - \eta_{mnil}^{AFM} L_m L_n - \eta_{mnil}^{\Phi P} \Phi_m \Phi_n + q_{mnil}^{(e)} u_{mn} \right) P_l + \beta_{Pij} P_i P_j^2 =$$
$$E_i^{ext} - E_i^r(x, \vec{r}) - \mu_{ji} M_j - \tilde{\mu}_{ji} L_j, \quad (S.3a)$$

$$\Gamma_M \frac{\partial L_i}{\partial t} + \left(\alpha_L \delta_{il} - \eta_{ilmn}^{AFM} P_m P_n - \eta_{ilmn}^{\Phi L}\Phi_m\Phi_n + q_{mnil}^{(l)} u_{mn}\right) L_l + \beta_{Lij} L_i L_j^2 = -\tilde{\mu}_{ij} P_j, \qquad (S.3b)$$

$$\Gamma_M \frac{\partial M_i}{\partial t} + \left(\alpha_M \delta_{il} - \eta_{ilmn}^{FM} P_m P_n - \eta_{ilmn}^{\Phi M}\Phi_m\Phi_n + q_{mnil}^{(m)} u_{mn}\right) M_l + \beta_{Mij} M_i M_j^2 = H_i^{ext} - H_i^r(x,\vec{r}) - \mu_{ij} P_j, \qquad (S.3c)$$

$$\Gamma_\Phi \frac{\partial \Phi_i}{\partial t} + \left(\alpha_\Phi \delta_{il} - \eta_{ilmn}^{\Phi P} P_m P_n - \eta_{ilmn}^{\Phi M} M_m M_n - \eta_{ilmn}^{\Phi L} L_m L_n + q_{mnil}^{(r)} u_{mn}\right) \Phi_l + \beta_{\Phi ij}\Phi_i\Phi_j^2 = 0. \qquad (S.3d)$$

Here $E_k^{ext}$ and $H_k^{ext}$ are external electric and magnetic fields, respectively.

In what follows we assume that the AFD long-range order significantly affects the FE, AFM and FM long-range orders, but not vice versa. Also, the FE long-range order significantly affects the AFM and FM long-range orders, but not vice versa, for the same reasons. The experimental results, obtained in this work, show that the FE and FM long-range orders are present for some content "x". To describe the x-dependence of the FE, AFM and FM long-range ordering we use the percolation theory approach can be used, meaning that the corresponding ordering appears above the percolation threshold, $x \geq x_{cr}^Z$, where the superscript Z = FE, FM or AFM, respectively.

Using the above arguments, the antiferromagnetic-paramagnetic (AFM-PM), ferromagnetic-paramagnetic (FM-PM) and ferroelectric-paraelectric (FE-PE) transition temperatures can be estimated as:

$$T_{PFY-Mn}^{FE} = T_Y^C - \frac{\eta_{\Phi P}(x)}{\alpha_{LT}}\Phi_S^2(0,x), \qquad (S.4a)$$

$$T_{PFY-Mn}^{AFM}(x) = T_Y^N \frac{(x-x_{cr}^{AFM})}{(1-x_{cr}^{AFM})} - \frac{\eta_{AFM}(x)}{\alpha_{LT}} P_S^2(0,x) - \frac{\eta_{\Phi L}(x)}{\alpha_{LT}}\Phi_S^2(0,x), \qquad (S.4b)$$

$$T_{PFY-Mn}^{FM}(x) = \theta_Y^C \frac{(x-x_{cr}^{FM})}{(1-x_{cr}^{FM})} - \frac{\eta_{FM}(x)}{\alpha_{MT}} P_S^2(0,x) - \frac{\eta_{\Phi M}(x)}{\alpha_{MT}}\Phi_S^2(0,x). \qquad (S.4c)$$

Hereinafter Y=N for PFN or Y=T for PFT; $T_{PFN}^C$ varies in the range (373 – 393) K for a bulk PFT and $T_{PFT}^C$ varies in the range (247 – 256) K for a bulk PFN; $T_Y^N$ varies in the range (143-170) K for a bulk PFT and (133-180) K for a bulk PFN, respectively. $P_S(0,x)$ and $\Phi_S(0,x)$ are the spontaneous polarization and AFD oxygen octahedra tilt at zero temperature, respectively. In Eqs.(S.4) the scalar approximation is used, and we regarded that the elastic stresses are absent.

Expressions (S.4) contain a relatively large number of fitting parameters, which should be related to the solution of LGD equations via the spontaneous polarization of the grains, $P_S(x, T)$. The solution in scalar approximation is

$$P_S(x,T) \approx \frac{P_S(x,0)}{1+\exp\left(\frac{T-T_C(x)}{\Delta_C(x,\omega)}\right)}, \quad (6)$$

where $P_S(x,0) \approx \sqrt{-\frac{\alpha_P - \eta_{FM} M_S^2 - \eta_{AFM} L_S^2 - \eta_{\Phi P} \Phi_S^2}{\beta_P}}$ is the zero-temperature spontaneous polarization considered in a scalar approximation for zero-disorder and strain-free conditions. The dispersion $\Delta_C(x, \omega)$ models the diffuseness of the polar order parameter transition to the ordered state.

# REFERENCES

[1] N. A. Spaldin, and R. Ramesh, Electric-field control of magnetism in complex oxide thin films. *MRS Bull.* 33, 1047 (2008); https://doi.org/10.1557/mrs2008.224

[2] M. Fiebig, T. Lottermoser, D. Meier, and M. Trassin. The evolution of multiferroics. *Nat. Rev. Mater* **1**, 16046 (2016); https://doi.org/10.1038/natrevmats.2016.46

[3] R. O. Cherifi, V. Ivanovskaya, L. C. Phillips, A. Zobelli, I. C. Infante, E. Jacquet, V. Garcia, S. Fusil, P. R. Briddon, N. Guiblin, A. Mougin, A. A. Ünal, F. Kronast, S. Valencia, B. Dkhil, A. Barthélémy, M. Bibes. Electric-field control of magnetic order above room temperature, *Nature Materials,* 13, 345–351 (2014); https://doi.org/10.1038/nmat3870

[4] M. Lilienblum, T. Lottermoser, S. Manz, S. M. Selbach, A. Cano, and M. Fiebig. Ferroelectricity in the multiferroic hexagonal manganites. *Nature Phys.* **11**, 1070–1073 (2015); https://doi.org/10.1038/nphys3468 .

[5] Y. Zemp, M Trassin, E Gradauskaite, B. Gao, S.-W. Cheong, T. Lottermoser, M. Fiebig, and M.s C. Weber. Magnetoelectric coupling in the multiferroic hybrid-improper ferroelectric $Ca_3Mn_{1.9}Ti_{0.1}O_7$. Physical Review B **109**, 184417 (2024); https://doi.org/10.1103/PhysRevB.109.184417

[6] D. A. Sanchez, N. Ortega, A. Kumar, R. Roque-Malherbe, R. Polanco, J. F. Scott, and R. S. Katiyar. Symmetries and multiferroic properties of novel room-temperature

magnetoelectrics: Lead iron tantalate–lead zirconate titanate (PFT/PZT). AIP Advances **1**, 042169 (2011); https://doi.org/10.1063/1.3670361

[7] L.W. Martin, R. Ramesh, Multiferroic and magnetoelectric heterostructures. Acta Materialia, **60**, 2449-2470 (2012); https://doi.org/10.1016/j.actamat.2011.12.024

[8] D.A. Sanchez, N. Ortega, A. Kumar, G. Sreenivasulu, R.S. Katiyar, J.F. Scott, D.M. Evans, M. Arredondo-Arechavala, A. Schilling, and J.M. Gregg. Room-temperature single phase multiferroic magnetoelectrics: Pb (Fe, M) x (Zr, Ti)(1− x) O3 [M= Ta, Nb]. J. Appl. Phys. **113**, 074105 (2013); https://doi.org/10.1063/1.4790317

[9] M.D. Glinchuk, R.O. Kuzian, Y.O. Zagorodniy, *et al.* Room-temperature ferroelectricity, superparamagnetism and large magnetoelectricity of solid solution $PbFe_{1/2}Ta_{1/2}O_3$ with $(PbMg_{1/3}Nb_{2/3}O_3)_{0.7}(PbTiO_3)_{0.3}$. *J Mater Sci* **55**, 1399–1413 (2020); https://doi.org/10.1007/s10853-019-04158-4

[10] S. A. Ivanov, R. Tellgren, H. Rundlof, N. W. Thomas, and S. Ananta. Investigation of the structure of the relaxor ferroelectric Pb (Fe1/2Nb1/2)O$_3$ by neutron powder diffraction. Journal of Physics: Condensed Matter **12**, 2393 (2000); https://doi.org/10.1088/0953-8984/12/11/305

[11] R.K. Mishra, R.N.P. Choudhary, and A. Banerjee. Bulk permittivity, low frequency relaxation and the magnetic properties of Pb (Fe1/2Nb1/2) O3 ceramics. Journal of Physics: Condensed Matter **22**, 025901 (2010); https://doi.org/10.1088/0953-8984/22/2/025901

[12] W. Kleemann, V. V. Shvartsman, P. Borisov, and A. Kania. Coexistence of antiferromagnetic and spin cluster glass order in the magnetoelectric relaxor multiferroic PbFe0.5Nb0.5O3. Phys. Rev. Lett. **105**, 257202 (2010); https://doi.org/10.1103/PhysRevLett.105.257202

[13] V.V. Bhat, A.M. Umarji, V.B. Shenoy, and U.V. Waghmare. Diffuse ferroelectric phase transitions in Pb-substituted PbFe1⁄2Nb1⁄2O3. Phys. Rev. B **72**, 014104 (2005); https://doi.org/10.1103/PhysRevB.72.014104

[14] C. Bharti, A. Dutta, S. Shannigrahi, T.P. Sinha. Electronic structure, magnetic and electrical properties of multiferroic PbFe1/2Ta1/2O3. Journal of Magnetism and Magnetic Materials **324**, 955 (2012); https://doi.org/10.1016/j.jmmm.2011.09.030

[15] N. Lampis, C. Franchini, G. Satta, A. Geddo-Lehmann, and S. Massidda. Electronic structure of PbFe$_{1/2}$Ta$_{1/2}$O$_3$: Crystallographic ordering and magnetic properties. Phys. Rev. B **69**, 064412 (2004); https://doi.org/10.1103/PhysRevB.69.064412

[16] W.Z. Zhu, A. Kholkin, P.Q. Mantas, J.L. Baptista, Preparation and characterisation of Pb (Fe 1/2 Ta 1/2) O 3 relaxor ferroelectric. J. Eur. Ceram. Soc. **20**, 2029–2034 (2000); https://doi.org/10.1016/S0955-2219(00)00094-7

[17] N. Kumar, A. Ghosh, R.N.P. Choudhary. Electrical behavior of Pb(Zr0.52Ti0.48)0.5(Fe0.5Nb0.5)0.5O3 ceramics. Mat. Chem. Phys. **130**, 381–386 (2011); https://doi.org/10.1016/j.matchemphys.2011.06.059

[18] M. D. Glinchuk, E. A. Eliseev, A. N. Morozovska, New room temperature multiferroics on the base of single-phase nanostructured perovskites. J. Appl. Phys. **116**, 054101 (2014); https://doi.org/10.1063/1.4891459

[19] M. D. Glinchuk, E. A. Eliseev, and A. N. Morozovska. Theoretical description of anomalous properties of novel room temperature multiferroics Pb(Fe1/2Ta1/2)x(Zr0.53Ti0.47)1-xO3 and Pb(Fe1/2Nb1/2)x(Zr0.53Ti0.47)1-xO3. *J. Appl. Phys*. **119**, 024102 (2016); https://doi.org/10.1063/1.4939584

[20] B.J. Rodriguez, S. Jesse, A.N. Morozovska, S.V. Svechnikov, D.A. Kiselev, A.L. Kholkin, A.A. Bokov, Z.-G. Ye, S.V. Kalinin, Real space mapping of polarization dynamics and hysteresis loop formation in relaxor-ferroelectric PbMg1/3Nb2/3O3–PbTiO3 solid solutions. J. Appl. Phys. **108**, 042006 (2010); https://doi.org/10.1063/1.3474961

[21] J. Petzelt, I. Rychetsky, D. Nuzhnyy, Dynamic ferroelectric–like softening due to the conduction in disordered and inhomogeneous systems: giant permittivity phenomena. Ferroelectrics, **426**, 171-193 (2012); https://doi.org/10.1080/00150193.2012.671732

[22] J. Petzelt, D. Nuzhnyy, V. Bovtun, M. Savinov, M. Kempa, I. Rychetsky, Broadband dielectric and conductivity spectroscopy of inhomogeneous and composite conductors. Phys. Stat. Sol. A **210**, 2259-2271 (2013); https://doi.org/10.1002/pssa.201329288

[23] O. S. Pylypchuk, S. E. Ivanchenko, M. Y. Yelisieiev, A. S. Nikolenko, V. I. Styopkin, B. Pokhylko, V. Kushnir, D. O. Stetsenko, O. Bereznykov, O. V. Leschenko, E. A. Eliseev, V. N. Poroshin, N. V. Morozovsky, V. V. Vainberg, and A. N. Morozovska.

Behavior of the Dielectric and Pyroelectric Responses of Ferroelectric Fine-Grained Ceramics. J. Amer. Ceram. Soc. **108**, e20391 (2025); https://doi.org/10.1111/jace.20391

[24] H. Han, Ch. Voisin, S. Guillemet-Fritsch, P. Dufour, Ch. Tenailleau, Ch. Turner, and J.C. Nino. Origin of colossal permittivity in $BaTiO_3$ via broadband dielectric spectroscopy. J. Appl. Phys. **113**, 024102 (2013); https://doi.org/10.1063/1.4774099

[25] L. Liu, S. Ren, J. Liu, F. Han, J. Zhang, B. Peng, D. Wang, A. A. Bokov, and Z.-G. Ye, Localized polarons and conductive charge carriers: Understanding $CaCu_3Ti_4O_{12}$ over a broad temperature range. Phys. Rev. B **99**, 094110 (2019); https://doi.org/10.1103/PhysRevB.99.094110

[26] I.A. Santos, and J.A. Eiras. Phenomenological description of the diffuse phase transition in ferroelectrics. J. Phys.: Condens. Matter **13**, 11733–11740 (2001); https://doi.org/10.1088/0953-8984/13/50/333

[27] Yu.O. Zagorodniy, R.O. Kuzian, I.V. Kondakova, M. Maryško, V. Chlan, H. Štěpánková, N.M. Olekhnovich et al. Chemical disorder and $^{207}Pb$ hyperfine fields in the magnetoelectric multiferroic $Pb(Fe_{1/2}Sb_{1/2})O_3$ and its solid solution with $Pb(Fe_{1/2}Nb_{1/2})O_3$. Phys. Rev. Mater. **2**, 014401 (2018); https://doi.org/10.1103/PhysRevMaterials.2.014401

[28] X. Li, Z. Hu, Y. Cho, X. Li, H. Sun, L. Cong, H.-J. Lin, S.-C. Liao, C.-T. Chen, A. Efimenko, C.J. Sahle, Y. Long, C. Jin, M.C. Downer, J.B. Goodenough, and J. Zhou. Charge Disproportionation and Complex Magnetism in a $PbMnO_3$ Perovskite Synthesized under High Pressure. Chem. Mater. **33**, 92−101 (2021); https://doi.org/10.1021/acs.chemmater.0c02706

[29] E. A. Eliseev, M. D. Glinchuk, V. Gopalan, A. N. Morozovska. Rotomagnetic couplings influence on the magnetic properties of antiferrodistortive antiferromagnets. J.Appl.Phys. **118**, 144101 (2015); https://doi.org/10.1063/1.4932211

[30] G. Catalan, J. Seidel, R. Ramesh, and J. F. Scott. Domain wall nanoelectronics. Rev. Mod. Phys. **84**, 119 (2012); https://doi.org/10.1103/RevModPhys.84.119

[31] D. V. Karpinsky, E. A. Eliseev, F. Xue, M. V. Silibin, A. Franz, M. D. Glinchuk, I. O. Troyanchuk, S. A. Gavrilov, V. Gopalan, L.-Q. Chen, and A. N. Morozovska.

Thermodynamic potential and phase diagram for multiferroic bismuth ferrite (BiFeO$_3$). npj Computational Materials **3**:20 (2017); https://doi.org/10.1038/s41524-017-0021-3

[32] B.I. Shklovskii and A.L. Efros. Electronic properties of doped semiconductors. (Springer-Verlag, Berlin 1984). 388 *Pages*.

[33] O. S. Pylypchuk, S. E. Ivanchenko, Y. O. Zagorodniy, M. E. Yelisieiev, O. V. Shyrokov, O. V. Leschenko, O. Bereznykov, D. Stetsenko, S. D. Skapin, E. A. Eliseev, V. N. Poroshin, V. V. Vainberg, and A. N. Morozovska. Relaxor-like Behavior of the Dielectric Response of Dense Ferroelectric Composites. Ceramics International **50,** 45465-45478 (2024); https://doi.org/10.1016/j.ceramint.2024.08.385

[34] O. S. Pylypchuk, V. V. Vainberg, V. N. Poroshin, O. V. Leshchenko, V. N. Pavlikov, I. V. Kondakova, S. E. Ivanchenko, L. P. Yurchenko, L. Demchenko, A. O. Diachenko, M. V. Karpets, M. P. Trubitsyn, E. A. Eliseev, and A. N. Morozovska. A colossal dielectric response of Hf$_x$Zr$_{1-x}$O$_2$ nanoparticles (2025); https://doi.org/10.48550/arXiv.2508.04697

[35] V. Martin-Mayor, J. Monforte-Garcia, A. Muñoz Sudupe, D. Navarro, G. Parisi, S. Perez-Gaviro, J. J. Ruiz-Lorenzo, S. F. Schifano, B. Seoane, A. Tarancon, R. Tripiccione, and D. Yllanes. Static versus Dynamic Heterogeneities in the *D* =3 Edwards-Anderson-Ising Spin Glass. Phys. Rev. Lett. **105**, 177202 (2010); https://doi.org/10.1103/PhysRevLett.105.177202

[36] R. Pirc, and R. Blinc. Spherical random-bond–random-field model of relaxor ferroelectrics. Physical Review B 60, 13470 (1999); https://doi.org/10.1103/PhysRevB.60.13470

[37] T.C. Choy. Effective Medium Theory. Oxford (UK): Clarendon Press; 1999. ISBN 978-0-19-851892-1.

[38] J. Petzelt, D. Nuzhnyy, V. Bovtun, D.A. Crandles. Origin of the colossal permittivity of (Nb+ In) co-doped rutile ceramics by wide-range dielectric spectroscopy. Phase Transitions **91**, 932 (2018). https://doi.org/10.1080/01411594.2018.1501801

[39] I. Rychetský, D. Nuzhnyy, J. Petzelt, Giant permittivity effects from the core–shell structure modeling of the dielectric spectra. Ferroelectrics **569**, 9 (2020). https://doi.org/10.1080/00150193.2020.1791659

[40] R. Simpkin. Derivation of Lichtenecker's logarithmic mixture formula from Maxwell's equations. IEEE Transactions on Microwave Theory and Techniques. **58**, 545 (2010); https://doi.org/10.1109/TMTT.2010.2040406.